\documentclass[a4paper,12pt]{article}
\usepackage{epsfig}
\usepackage{color}
\usepackage[numbers,sort&compress]{natbib}

\newlength{\dinwidth}
\newlength{\dinmargin}
\begin{document}
\title{Revisiting  $B_s\to \mu^+\mu^-$
and $B\to K^{(*)}\mu^+\mu^-$ decays in the MSSM with and without
R-parity}

\author{ Ru-Min Wang$^{1}$\thanks{E-mail: ruminwang@gmail.com},
~~Yuan-Guo Xu$^{1}$,
~~Yi-Long Wang$^{1}$,
~~Ya-Dong Yang$^{2,3}$
  \\
{\scriptsize {$^1$ \it College of Physics and Electronic
Engineering, Xinyang Normal University,
 Xinyang, Henan 464000, China
}}
\\
 {\scriptsize  {$^2$ \it Institute of Particle Physics, Huazhong Normal University, Wuhan,
Hubei 430079, P. R. China }}
\\
 {\scriptsize  {$^3$ \it Key Laboratory of Quark and Lepton Physics,
Ministry of Education, P.R. China }}
 }
 \maketitle
\vspace{-0.4cm}
\begin{abstract}

The rare decays $B_s\to \mu^+\mu^-$ and $B\to K^{(*)}\mu^+\mu^-$ are
sensitive to new particles and couplings via their interferences
with the standard model contributions. Recently, the upper bound on
$\mathcal{B}(B_s\to \mu^+\mu^-)$ has been improved significantly by
the CMS, LHCb, CDF, and D{\O} experiments.  Combining with the
measurements of $\mathcal{B}(B\to K^{(*)}\mu^+\mu^-)$, we derive
constraints on the relevant parameters of minimal supersymmetic
standard model with and without R-parity, and examine their
contributions to the dimuon forward-backward asymmetry in $B\to
K^{*}\mu^+\mu^-$ decay.  We find that (i) the contribution of
R-parity violating coupling products
$\lambda^{\prime}_{2i2}\lambda^{\prime*}_{2i3}$ due to squark
exchange is comparable with the theoretical uncertainties in
 $B\to K \mu^+\mu^-$ decay, but still could be significant in
 $B\to K^{*}\mu^+\mu^-$ decay and could account for the forward-backward asymmetry
 in all dimuon invariant mass regions; (ii) the constrained  mass
 insertion $(\delta^{u}_{LL})_{23}$
 could have significant  contribution to  $d \mathcal{A}_{FB}(B\to
  K^{*}\mu^+\mu^-)/ds$, and such effects are
 favored  by thr recent results of the Belle, CDF, and LHCb experiments.

\end{abstract}


\newpage
\section{Introduction}
Recently, using the 7$fb^{-1}$ data set,  the CDF Collaboration at the
Fermilab Tevatron  has observed an excess of $B_{s}$ candidates
\cite{Aaltonen:2011fi}, which is compatible with
\begin{eqnarray}
\mathcal{B}(B_s\to \mu^+\mu^-)=(1.8^{+1.1}_{-0.9})\times10^{-8},
\end{eqnarray}
and provided the corresponding upper limit of $\mathcal{B}(B_s\to
\mu^+\mu^-)<4.0\times10^{-8}$ at 95\% confidence level (CL).

At the same time, searches for $B_s\to \mu^+\mu^-$ have also been
made by the CMS and LHCb Collaborations
\cite{Chatrchyan:2011kr,arXiv:1112.1600,LHCBdata}, respectively, at
the Large Hadron Collider at CERN.  The combined results of the
searches by the CMS and LHCb Collaborations  in the upper limits
\cite{CMSLHCB} are
  \begin{eqnarray}
   \mathcal{B}(B_s\to \mu^+\mu^-)<1.08\times10^{-8}~ \mbox{at}~ 95\%~ \rm{CL},\\
    \mathcal{B}(B_s\to \mu^+\mu^-)<0.90\times10^{-8}~ \mbox{at}~ 90\% ~\rm{CL},
\end{eqnarray}
which have improved the previous upper bounds \cite{D0-CDF}
significantly.

$B_s\to \mu^+\mu^-$  decay is a known sensitive probe to the
presence of  new physics (NP). In the standard model (SM), it occurs
via penguin or box diagrams and is strongly  helicity suppressed.
Its SM prediction is $(3.2\pm0.2)\times10^{-9}$
 \cite{Buras:2010mh}.  Generally, NP could enhance the $B_s\to \mu^+\mu^-$
decay rate very much, and thus the upper bound of
$\mathcal{B}(B_s\to \mu^+\mu^-)$ is taken as a strong constraint
when  a NP model is discussed. As a cross-check, one usually
needs to investigate the semileptonic rare decays $B\to K \mu^+
\mu^-$ and $B\to K^{*} \mu^+ \mu^-$    which are also governed by
the flavor changing neutral current transition $b\to s \mu^+
\mu^-$ but not helicity suppressed.  Many observables of $B \to
K^{(*)}\mu^+\mu^-$ have been observed by several experiments: BABAR
\cite{:2008ju}, Belle  \cite{:2009zv}, CDF \cite{Aaltonen:2011ja},
and LHCb \cite{Aaij:2011aa}. As many of them agree with the SM
predictions within their error bars, however, the dimuon
forward-backward asymmetry of $B\to K^{*} \mu^+ \mu^-$ at the low
 region of the dimuon invariant mass  is not consistently measured
by Belle \cite{:2009zv}, CDF \cite{Aaltonen:2011ja}, and LHCb
\cite{Aaij:2011aa}.

Any NP that alters $\mathcal{B}(B_s\to \mu^+\mu^-)$ would
necessarily alter observables in $B\to K^{(*)} \mu^+ \mu^-$ decays;
examples of the latter are the differential branching ratio and
forward-backward asymmetry.
 The NP
effects in the $b\to s\mu^+\mu^-$ flavor changing neutral current transition have been extensively
investigated, for instance, in Refs.
 \cite{Palle:2011mk,Beskidt:2011qf,Akeroyd:2011kd,Altmannshofer:2011rm,Alok:2010zd,Lunghi:2010tr,Chang:2010zy,
Altmannshofer:2009ne,Altmannshofer:2008dz,Alok:2009wk}.
 In this paper,  following
closely the analysis of Ref.  \cite{Xu:2006vk}, we will update the
constraints on the R-parity violating (RPV) minimal supersymmetric
standard model (MSSM) in light of the new experimental data on
$B_s\to \mu^+\mu^- $ and $B \to K^{(*)}\mu^+\mu^-$. Additionally, we
will extend our analysis to the R-parity conserving (RPC)  MSSM
scenario with the mass insertion (MI) approximation
\cite{hep-ph/9604387,DFPD-88-TH-8}. Using a combination of the
limits of $\mathcal{B}(B_s\to \mu^+\mu^-)$ from CDF, LHCb and CMS
 \cite{Aaltonen:2011fi,CMSLHCB} as well as the experimental bounds of
$\mathcal{B}(B\to K^{(*)}\mu^+\mu^-)$  \cite{PDG}, we will obtain
the new limits on the relevant supersymmetric coupling parameters.
Then we will use the constrained parameter spaces to examine
their effects on some observables in these decays, especially
$d\mathcal{A}_{FB}(B\to K^{*}\mu^{+}\mu^{-})/ds$.

The paper is arranged as follows. In Sec. 2,  we present a very
brief theoretical introduction to $B_s\to \mu^+\mu^-$ and $B\to
K^{(*)}\mu^+\mu^-$  processes. In Sec. 3, we deal with the
numerical results. We display the constraints implied by the new
experimental data on the RPV and RPC MSSM parameter spaces and
discuss the implications for the $B \to K^{(*)}\mu^+\mu^-$ invariant
mass spectra and forward-backward asymmetries.
 Section 4 contains our  conclusion.

\section{The theoretical framework for $B_s\to \mu^+\mu^- $ and $B \to K^{(*)}\mu^+\mu^-$ decays}
\label{hha}

\subsection{The leptonic decay $B_s\to \mu^+\mu^-$ }

The branching ratio for $B_s\to \mu^+\mu^-$ can be written as
 \cite{Bobeth:2001sq,Altmannshofer:2009ne}
\begin{eqnarray}
\mathcal{B}(B_s\to
\mu^+\mu^-)=\frac{\tau_{B_s}m^3_{B_s}f^2_{B_s}}{32\pi}\sqrt{1-\frac{4m^2_{\mu}}{m^2_{B_s}}}\left[
\left|F_B\right|^2 \left(1-\frac{4m^2_{\mu}}{m^2_{B_s}}
\right)+\left|F_A\right|^2 \right],\label{eq.pureleptonBr}
\end{eqnarray}
where
\begin{eqnarray}
F_A&=&\frac{2m_\mu}{m_{B_s}}\left(C_A-\widetilde{C}_A\right)+m_{B_s}\left(C_P-\widetilde{C}_P\right),\nonumber\\
F_B&=&m_{B_s}\left(C_S-\widetilde{C}_S\right).
\end{eqnarray}

The SM result for the branching ratio may be obtained from Eq.
(\ref{eq.pureleptonBr}) by setting
$\widetilde{C}_A=C_S=\widetilde{C}_S=C_P=\widetilde{C}_P=0$ and
\begin{eqnarray}
C_A=\frac{G_F\alpha_e}{\sqrt{2}\pi
\mbox{sin}^2\theta_W}V_{tb}V^*_{ts}Y(x_t).
\end{eqnarray}

In the MSSM without R-parity, the branching ratio may be obtained by
setting  \cite{Xu:2006vk}
\begin{eqnarray}
C'_A&=&-\frac{\lambda'_{2i2}\lambda_{2i3}'^{*}}{4m^2_{\tilde{u}_{iL}}},\nonumber\\
C_S&=&-C_P~=~-\frac{\lambda_{i22}\lambda_{i23}'^{*}}{4m_bm^2_{\tilde{\nu}_{iL}}},\nonumber\\
C'_S&=&C'_P~=~-\frac{\lambda^*_{i22}\lambda'_{i32}}{4m_bm^2_{\tilde{\nu}_{iL}}}.
\end{eqnarray}

In the MSSM with R-parity, the branching ratio can obtained by using
the expressions $C_S,\widetilde{C}_S,C_P$ and $\widetilde{C}_P$ can
be found in Ref.  \cite{Altmannshofer:2009ne}; and
$\widetilde{C}_A=0$ in this case.

\subsection{The semileptonic decays $B\to K^{(*)} \mu^+\mu^-$}
In the SM, the double differential decay branching ratios
$\frac{d^2\mathcal{B}^K}{d\hat{s}d\hat{u}}$ and
$\frac{d^2\mathcal{B}^{K^*}}{d\hat{s}d\hat{u}}$ for the decays $B\to
K\mu^+\mu^-$ and $B\to K^*\mu^+\mu^-$, respectively, may be written
as   \cite{Ali:1999mm}
\begin{eqnarray}
\frac{d^2\mathcal{B}^{K}_{SM}}{d\hat{s}d\hat{u}}&=&\tau_B
\frac{G^2_F\alpha_{e}^2m_B^5}{2^{11}\pi^5}|V^*_{ts}V_{tb}|^2 \nonumber\\
&&\times\Bigg\{(|A'|^2+|C'|^2)(\lambda-\hat{u}^2)\nonumber\\
&&+|C'|^24\hat{m}^2_\mu(2+2\hat{m}^2_K-\hat{s})
+Re(C'D'^*)8\hat{m}^2_\mu(1-\hat{m}^2_K)+|D'|^24\hat{m}^2_\mu\hat{s}\Bigg\},\label{BK}\\
\frac{d^2\mathcal{B}^{K^*}_{SM}}{d\hat{s}d\hat{u}}
&=&\tau_B\frac{G^2_F\alpha_{e}^2m_B^5}{2^{11}\pi^5}|V^*_{ts}V_{tb}|^2\nonumber\\
&&\times\left\{\frac{|A|^2}{4}\Big(\hat{s}(\lambda+\hat{u}^2)+4\hat{m}^2_\mu\lambda\Big)
+\frac{|E|^2}{4}\Big(\hat{s}(\lambda+\hat{u}^2)
-4\hat{m}^2_\mu\lambda\Big)\right.\nonumber\\
&&+\frac{1}{4\hat{m}^2_{K^*}}\Big[|B|^2\Big(\lambda-\hat{u}^2
+8\hat{m}^2_{K^*}(\hat{s}+2\hat{m}^2_\mu)\Big)
+|F|^2\Big(\lambda-\hat{u}^2+8\hat{m}^2_{K^*}(\hat{s}-4\hat{m}^2_\mu)\Big)\Big]\nonumber\\
&&-2\hat{s}\hat{u}\Big[Re(BE^*)+Re(AF^*)\Big]\nonumber\\
&&+\frac{\lambda}{4\hat{m}^2_{K^*}}\Big[|C|^2(\lambda-\hat{u}^2)+
|G|^2(\lambda-\hat{u}^2+4\hat{m}^2_\mu(2+2\hat{m}^2_{K^*}-\hat{s})\Big)\Big]\nonumber\\
&&-\frac{1}{2\hat{m}^2_{K^*}}\Big[Re(BC^*)(1-\hat{m}^2_{K^*}-\hat{s})(\lambda-\hat{u}^2)\nonumber\\
&&~~~~~~~~~+Re(FG^*)\Big((1-\hat{m}^2_{K^*}
-\hat{s})(\lambda-\hat{u}^2)+4\hat{m}^2_\mu\lambda\Big)\Big]\nonumber\\
&&\left.-2\frac{\hat{m}^2_\mu}{\hat{m}^2_{K^*}}
\lambda\Big[Re(FH^*)-Re(GH^*)(1-\hat{m}^2_{K^*})\Big]
+|H|^2\frac{\hat{m}^2_\mu}{\hat{m}^2_{K^*}}\hat{s}\lambda \right\},
\label{BKs}
\end{eqnarray}
where $ p = p_B+p_{K^{(*)}}$, $s = q^2$, and $q = p_++p_-$ ($p_\pm$
the four-momenta of the muons), and the auxiliary functions $A-H$
can be found in Ref.  \cite{Ali:1999mm}. The hat denotes
normalization in terms of the B-meson mass, $m_B$, e.g.,
$\hat{s}=s/m_B^2$, $\hat{m}_q=m_q/m_B$.

In the MSSM without R-parity, the double differential decay
branching ratios including the squark exchange contributions could be gotten
from Eqs. (\ref{BK}-\ref{BKs})  by the replacements
 \cite{Xu:2006vk}
\begin{eqnarray}
A'(\hat{s})&\rightarrow&A'(\hat{s})+\frac{f_+^{B\to K}(\hat{s})}{W}\sum_i\frac{\lambda'_{2i2}\lambda_{2i3}'^{*}}{8m^2_{\tilde{u}_{iL}}},\nonumber\\
C'(\hat{s})&\rightarrow&C'(\hat{s})-\frac{f_+^{B\to K}(\hat{s})}{W}\sum_i\frac{\lambda'_{2i2}\lambda_{2i3}'^{*}}{8m^2_{\tilde{u}_{iL}}},\nonumber\\
A(\hat{s})&\rightarrow&A(\hat{s})+\frac{1}{W} \left[\frac{2V^{B\to
K^*}(\hat{s})}{m_B+m_{K^*}}m^2_B\right]\sum_i\frac{\lambda'_{2i2}\lambda_{2i3}'^{*}}{8m^2_{\tilde{u}_{iL}}},\nonumber\\
B(\hat{s})&\rightarrow&B(\hat{s})+\frac{1}{W} \left[-(m_B+m_{K^*})A_1^{B\to K^*}(\hat{s})\right]
\sum_i\frac{\lambda'_{2i2}\lambda_{2i3}'^{*}}{8m^2_{\tilde{u}_{iL}}},\nonumber\\
C(\hat{s})&\rightarrow&C(\hat{s})+\frac{1}{W} \left[\frac{A_2^{B\to K^*}(\hat{s})}{m_B+m_{K^*}}m_B^2\right]
\sum_i\frac{\lambda'_{2i2}\lambda_{2i3}'^{*}}{8m^2_{\tilde{u}_{iL}}},\nonumber\\
D(\hat{s})&\rightarrow&D(\hat{s})+\frac{1}{W}\left[\frac{~2m_{K^*}}{\hat{s}}\Big(A_3^{B\to
K^*}(\hat{s})-A_0^{B\to K^*}(\hat{s})\Big)\right]
\sum_i\frac{\lambda'_{2i2}\lambda_{2i3}'^{*}}{8m^2_{\tilde{u}_{iL}}},\nonumber\\
E(\hat{s})&\rightarrow&E(\hat{s})-\frac{1}{W} \left[\frac{2V^{B\to
K^*}(\hat{s})}{m_B+m_{K^*}}m^2_B\right]\sum_i\frac{\lambda'_{2i2}\lambda_{2i3}'^{*}}{8m^2_{\tilde{u}_{iL}}},\nonumber\\
F(\hat{s})&\rightarrow&F(\hat{s})-\frac{1}{W}
\left[-(m_B+m_{K^*})A_1^{B\to K^*}(\hat{s})\right]
\sum_i\frac{\lambda'_{2i2}\lambda_{2i3}'^{*}}{8m^2_{\tilde{u}_{iL}}},\nonumber\\
G(\hat{s})&\rightarrow&G(\hat{s})-\frac{1}{W} \left[\frac{A_2^{B\to
K^*}(\hat{s})}{m_B+m_{K^*}}m_B^2\right]
\sum_i\frac{\lambda'_{2i2}\lambda_{2i3}'^{*}}{8m^2_{\tilde{u}_{iL}}},\nonumber\\
H(\hat{s})&\rightarrow&H(\hat{s})-\frac{1}{W}\left[\frac{~2m_{K^*}}{\hat{s}}\Big(A_3^{B\to
K^*}(\hat{s})-A_0^{B\to K^*}(\hat{s})\Big)\right]
\sum_i\frac{\lambda'_{2i2}\lambda_{2i3}'^{*}}{8m^2_{\tilde{u}_{iL}}},
\end{eqnarray}
where $W=-\frac{G_F\alpha_{e}}{2\sqrt{2}~\pi}V^*_{ts}V_{tb}m_B$.

The sneutrino exchange contributions are summarized as
\begin{eqnarray}
\frac{d^2\mathcal{B}^{K}_{\tilde{\nu}}}{d\hat{s}d\hat{u}}
&=&\tau_B\frac{m^3_B}{2^7\pi^3}\left\{\frac{}{}\right.
Re(WA'\mathcal{T}'^{*}_{S})(2\hat{m}_{\mu}\hat{u})
+Re(WC'\mathcal{T}'^{*}_{P})(1-\hat{m}_K^2)(-2\hat{m}_{\mu})\nonumber\\
&&+Re(WD'\mathcal{T}'^{*}_{P})(-2\hat{m}_{\mu}\hat{s})
+|\mathcal{T}'_{S}|^2(\hat{s}-2\hat{m}^2_{\mu})\left.\frac{}{}\right\},\\
\frac{d^2\mathcal{B}^{K^*}_{\tilde{\nu}}}{d\hat{s}d\hat{u}}&=&\tau_B\frac{m^3_B}{2^7\pi^3}
\Bigg\{-\frac{\hat{m}^2_{\mu}}{\hat{m}^2_{K^*}}
\Bigg[Im(WB\mathcal{T}^{*}_{S})
\Big(\lambda^{-\frac{1}{2}}\hat{u}(1-\hat{m}^2_{K^*}-\hat{s})\Big)\nonumber\\
&&+Im(WC\mathcal{T}^{*}_{S})
\lambda^{\frac{1}{2}}\hat{u}-Im(WF\mathcal{T}^{*}_{P})\lambda^{\frac{1}{2}}\nonumber\\
&&+Im(WG\mathcal{T}^{*}_{P})\lambda^{\frac{1}{2}}(1-\hat{m}^2_{K^*})\Bigg]
+|\mathcal{T}_{S}|^2(\hat{s}-2\hat{m}^2_{\mu})\Bigg\},
\end{eqnarray}
with
\begin{eqnarray}
\mathcal{T}'_S=f_+^{B\to K}(\hat{s})\frac{m^2_B-m^2_K}{\overline{m}_b-\overline{m}_s}
\sum_i\left(\frac{\lambda^{*}_{i22}\lambda_{i32}'}{8m^2_{\tilde{\nu}_{iL}}}+\frac{\lambda_{i22}\lambda_{i23}'^{*}}{8m^2_{\tilde{\nu}_{iL}}}\right),\nonumber\\
\mathcal{T}'_P=f_+^{B\to
K}(\hat{s})\frac{m^2_B-m^2_K}{\overline{m}_b-\overline{m}_s}
\sum_i\left(\frac{\lambda^{*}_{i22}\lambda_{i32}'}{8m^2_{\tilde{\nu}_{iL}}}-\frac{\lambda_{i22}\lambda_{i23}'^{*}}{8m^2_{\tilde{\nu}_{iL}}}\right),\nonumber\\
\mathcal{T}_S=\left[\frac{i}{2}\frac{A_0^{B\to
K^*}(\hat{s})}{\overline{m}_b+\overline{m}_s}
\lambda^{\frac{1}{2}}m^2_B\right]
\sum_i\left(\frac{\lambda^{*}_{i22}\lambda_{i32}'}{8m^2_{\tilde{\nu}_{iL}}}-\frac{\lambda_{i22}\lambda_{i23}'^{*}}{8m^2_{\tilde{\nu}_{iL}}}\right),\nonumber\\
\mathcal{T}_P=\left[\frac{i}{2}\frac{A_0^{B\to
K^*}(\hat{s})}{\overline{m}_b+\overline{m}_s}\lambda^{\frac{1}{2}}m^2_B\right]
\sum_i\left(\frac{\lambda^{*}_{i22}\lambda_{i32}'}{8m^2_{\tilde{\nu}_{iL}}}+\frac{\lambda_{i22}\lambda_{i23}'^{*}}{8m^2_{\tilde{\nu}_{iL}}}\right).
\end{eqnarray}

In the MSSM with R-parity, all the effects arise from the RPC  MIs
contributing to $C_7,\widetilde{C}^{eff}_9,\widetilde{C}_{10}$, and
they are
\begin{eqnarray}
C^{RPC}_7&=&C^{Diag}_7+C^{MI}_7+nC'^{MI}_7,\nonumber\\
 (C^{eff}_{9})^{RPC}&=& (\widetilde{C}^{eff}_{9})^{Diag}+ (\widetilde{C}^{eff}_{9})^{MI}+ n(C'^{eff}_{9})^{MI},\nonumber\\
C^{RPC}_{10}&=&\widetilde{C}^{Diag}_{10}+\widetilde{C}^{MI}_{10}+nC'^{MI}_{10},
\end{eqnarray}
where $n=1$ for  decay $B\to K \mu^+\mu^-$  as well as for the terms
related to the form factors $V$ and $T_1$ in $B\to K^* \mu^+\mu^-$
decay, $n=-1$ for the terms related to the form factors
$A_0,A_1,A_2,T_2$ and $T_3$ in $B\to K^* \mu^+\mu^-$ decay.
$C_7^{Diag,MI},(\widetilde{C}^{eff}_{9})^{Diag,MI}$,
$\widetilde{C}^{Diag,MI}_{10}$, $C'^{MI}_7$, $(C'^{eff}_{9})^{MI}$
and $C'^{MI}_{10}$ have been estimated in Refs.
 \cite{Lunghi:1999uk,Cho:1996we,Hewett:1996ct}.  The results for $\mathcal{B}^K$ and
$\mathcal{B}^{K^*}$ including MI effects can be obtained from Eqs.
(\ref{BK}-\ref{BKs}) by the following replacements
 \cite{Altmannshofer:2008dz,Lunghi:2010tr}:
\begin{eqnarray}
C^{SM}_7&\rightarrow&C^{SM}_7+C^{RPC}_7,\nonumber\\
(C^{eff}_{9})^{SM}&\rightarrow& (C^{eff}_{9})^{SM}+ (C^{eff}_{9})^{RPC},\nonumber\\
C^{SM}_{10}&\rightarrow&C^{SM}_{10}+C^{RPC}_{10}.
\end{eqnarray}

From  the total double differential branching ratios,  we can get
the dimuon  forward-backward asymmetries
 \cite{Ali:1999mm}
\begin{eqnarray}
\mathcal{A}_{FB}(B\to K^{(*)}\mu^+\mu^-)=\int
d\hat{s}~\frac{\int^{+1}_{-1}\frac{d^2\mathcal{B}(B\to
K^{(*)}\mu^+\mu^-)}{d\hat{s}dcos\theta}sign(cos\theta)dcos\theta}
{\int^{+1}_{-1}\frac{d^2\mathcal{B}(B\to
K^{(*)}\mu^+\mu^-)}{d\hat{s}dcos\theta}dcos\theta}.
\end{eqnarray}

\section{Numerical results and analyses}
We will present our numerical results and analysis in this section.
 When we study the effects due to MSSM with and without R-parity,
 we consider only one new coupling at one time, neglecting the interferences between
different new couplings, but keeping their interferences
with the SM amplitude. The input parameters are collected in the
Appendix, and the following experimental data will be used to
constrain parameters of  the relevant new couplings
 \cite{CMSLHCB,PDG}:
\begin{eqnarray}
&&\mathcal{B}(B_s\to \mu^+\mu^-)<1.08\times10^{-8}  ~(\mbox{at } 95\%~\mbox{CL}),\nonumber\\
&&\mathcal{B}(B \to K\mu^+\mu^-)=(0.48\pm0.06)\times10^{-6}, \nonumber\\
&& \mathcal{B}(B \to
K^*\mu^+\mu^-)=(1.15\pm0.15)\times10^{-6}.\label{Eq:exp}
\end{eqnarray}
To be conservative, we use the input parameters  varied randomly
within 1$\sigma$ variance and the experimental bounds at 95\% CL. We
do not impose the experimental bounds from $d\mathcal{A}_{FB}(B \to
K^*\mu^+\mu^-)/ds$ and leave it as predictions of the restricted
parameter spaces of the two NP scenarios, and compare them with the
experimental results in Refs.
\cite{:2009zv,Aaltonen:2011ja,Aaij:2011aa}.

\subsection{RPV MSSM effects}

First, we will  consider the RPV effects and  further constrain
the relevant RPV couplings from  the new experimental data of
$\mathcal{B}(B_s\to \mu^+\mu^-)$ and $\mathcal{B}(B \to
K^{(*)}\mu^+\mu^-)$ given in Eq. (\ref{Eq:exp}). As given  in Sec.
\ref{hha}, there are three   RPV coupling products, which are
$\lambda'_{2i2}\lambda'^*_{2i3}$ due to squark exchange as well as
$\lambda_{i22}\lambda'^*_{i23}$ and $\lambda^*_{i22}\lambda'_{i32}$
due to sneutrino exchange, relevant  to $B_s\to \mu^+\mu^-$ and
$B\to K^{(*)}\mu^+\mu^-$ decays.

\begin{figure}[b]
\begin{center}
\includegraphics[scale=1.3]{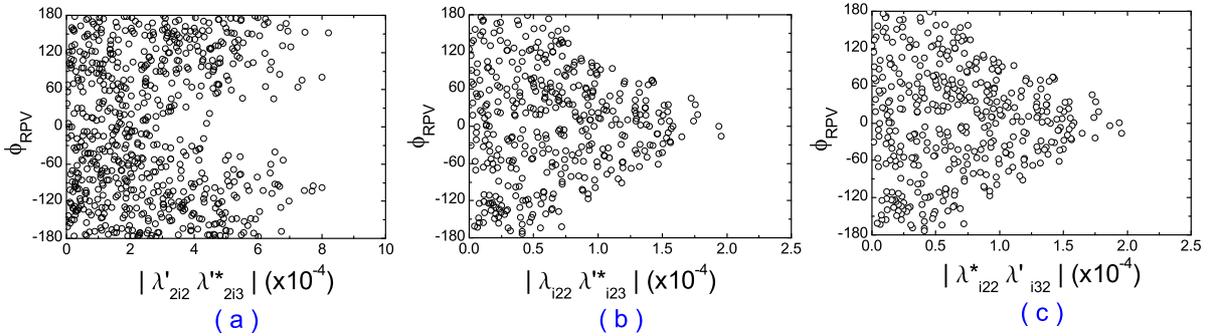}
\end{center}
\vspace{-0.4cm}
 \caption{ The allowed RPV parameter spaces  with 500 GeV sfermions,
 and the RPV weak phase $(\phi_{RPV})$ is given in degree.}
 \label{fig:bounds}
\end{figure}

Our new bounds for three RPV coupling products from the 95\% CL
experimental data are demonstrated in Fig. \ref{fig:bounds}.
\begin{table}[h]
\caption{Bounds for  the relevant RPV coupling products by $B \to
K^{(*)}\mu^+\mu^-$ and $B_s \to \mu^+\mu^-$ decays for 500 GeV
sfermions, and previous bounds are listed for comparison.
}\label{Tab:bounds}
\begin{center}
\begin{tabular}{c|l|l}\hline\hline
Couplings&~~~~Bounds & Previous bounds  \cite{Xu:2006vk}\\
\hline
$|\lambda'_{2i2}\lambda'^*_{2i3}| $&$\leq8.2\times 10^{-4}$&$\leq11.5\times 10^{-4}$ \\
$|\lambda_{i22}\lambda'^*_{i32}| $&$\leq2.0\times 10^{-4}$&$\leq4.5\times 10^{-4}$ \\
$|\lambda^*_{i22}\lambda'_{i23}| $&$\leq2.0\times
10^{-4}$&$\leq4.3\times 10^{-4}$ \\\hline\hline
\end{tabular}
\end{center}
\end{table}
And the upper limits for  the relevant RPV coupling products  by
$\mathcal{B}(B \to K^{(*)}\mu^+\mu^-)$ and $\mathcal{B}(B _s\to
\mu^+\mu^-)$  are summarized  in Table \ref{Tab:bounds}.   For
comparison,  our previous bounds on these quadric coupling products
are also listed. From Fig. \ref{fig:bounds} and Table
\ref{Tab:bounds}, one can find that  all three RPV coupling products
are restricted, and the upper limits of
$|\lambda_{i22}\lambda'^*_{i32}|$ and
$|\lambda^*_{i22}\lambda'_{i23}|$ are improved by about a factor of
2 by the new experimental data. Notice that we assume the masses of
sfermions are 500 GeV. For other values of the sfermion masses, the
bounds on the couplings in this paper can be easily obtained by
scaling them by factor of
$\tilde{f}^2\equiv(\frac{m_{\tilde{f}}}{500GeV})^2$.

\begin{figure}[b]
\begin{center}
\includegraphics[scale=0.5]{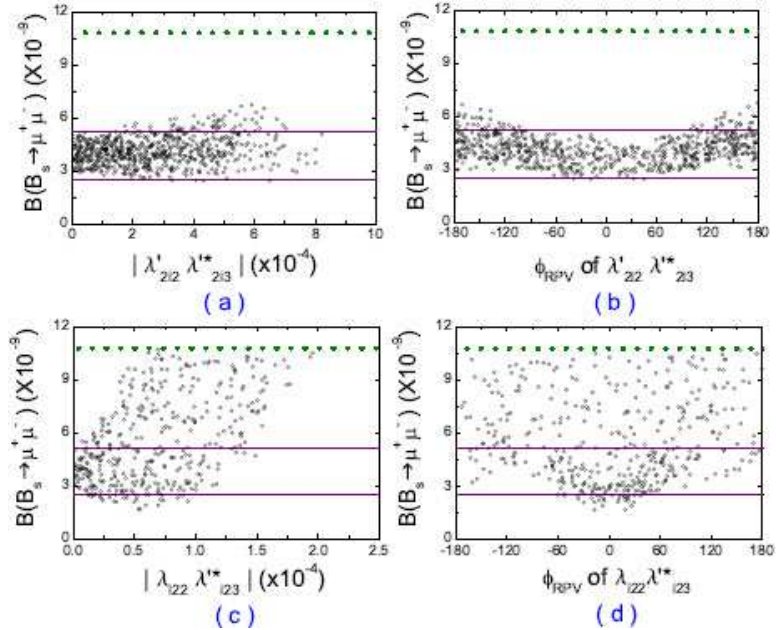}
\end{center}
\vspace{-0.4cm} \caption{ The constrained  RPV coupling effects on
$\mathcal{B}(B_s\to \mu^+\mu^-)$. The olive (violet) horizontal
dotted (solid) lines denote the limits of the 95\% CL measurements (SM
predictions).}\label{fig:Bsmummu}
\end{figure}

Now we will analyze the constrained RPV effects on
$\mathcal{B}(B_s\to \mu^+\mu^-)$.  The sensitivities of
$\mathcal{B}(B_s\to \mu^+\mu^-)$ to the constrained RPV couplings
are shown in Fig. \ref{fig:Bsmummu}.
The limits of the measurements  at 95\% CL and the SM predictions
with $1\sigma$ theoretical uncertainties are also displayed in Fig.
\ref{fig:Bsmummu}  for convenient comparison. Figs.
\ref{fig:Bsmummu} (a) and (b) show the constrained effects of the
modulus and weak phase of  t-channel squark exchange coupling
$\lambda'_{2i2}\lambda'^*_{2i3}$, respectively. As shown in Figs.
\ref{fig:Bsmummu} (a-b), with the contribution of
$\lambda'_{2i2}\lambda'^*_{2i3}$ included, $\mathcal{B}(B_s\to
\mu^+\mu^-)$ is lower than its experimental upper limit
\cite{CMSLHCB}. Besides the constraints from $\mathcal{B}(B_s\to
K^{(*)}  \mu^+\mu^-)$,
  $\lambda'_{2i2}\lambda'^*_{2i3}$ coupling is
not further constrained by the new experimental upper limit from CMS
and LHCb since its contribution to $\mathcal{B}(B_s\to \mu^+\mu^-)$
is suppressed by $m^2_{\mu}/m^2_B$.
Additionally, the allowed parameter space of
$\lambda'_{2i3}\lambda'^*_{2i2}$ would be excluded if  the 68\% CL
experimental determination $\mathcal{B}(B_s\to
\mu^+\mu^-)=(1.8^{+1.1}_{-0.9})\times10^{-8}$ \cite{Aaltonen:2011fi}
by the CDF Collaboration were taken as a constraint.
Two s-channel sneutrino exchange  contributions to
$\mathcal{B}(B_s\to \mu^+\mu^-)$ are very similar to each other. We
would take the $\lambda_{i22}\lambda'^*_{i23}$ contribution as an
example, which is shown by  Figs. \ref{fig:Bsmummu} (c-d). We can
see that $\mathcal{B}(B_s\to \mu^+\mu^-)$ is sensitive to both the
modulus and phase of $\lambda_{i22}\lambda'^*_{i23}$, and
$\mathcal{B}(B_s\to \mu^+\mu^-)$  not only  could be increased but also could
be decreased by  the presence of $\lambda_{i22}\lambda'^*_{i23}$
coupling.
Generally,
the $\lambda_{i22}\lambda'^*_{i23}$ coupling could alter
$\mathcal{B}(B_s\to \mu^+\mu^-)$ significantly
 since  its contribution  is not helicity suppressed
by $m^2_{\mu}/m^2_B$. Thus, the constraint   on
$\lambda_{i22}\lambda'^*_{i23}$ is due to the bound of
$\mathcal{B}(B_s\to \mu^+\mu^-)$ \cite{CMSLHCB}.

Then  we turn to analyzing the constrained RPV effects in  $B\to
K^{(*)}\mu^+\mu^-$ decays. Using the new constrained parameter
spaces shown in Fig. \ref{fig:bounds}, we will give the RPV effects
on the dimuon invariant mass spectra and the forward-backward
asymmetries of $B\to K^{(*)}\mu^+\mu^-$ decays.
\begin{figure}[th]
\begin{center}
\includegraphics[scale=0.78]{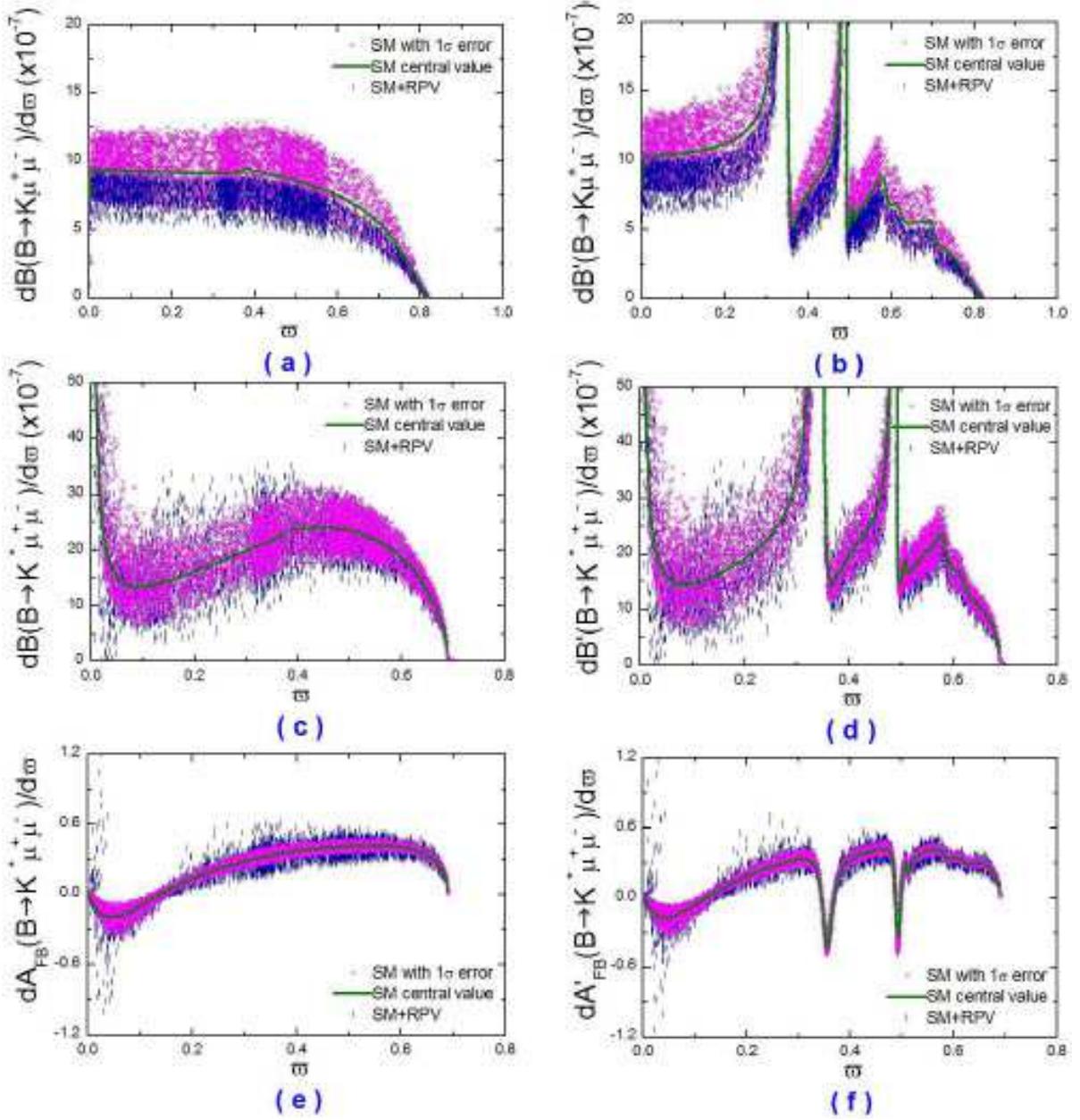}
\end{center}
\vspace{-0.4cm} \caption{ The effects of RPV coupling
$\lambda'_{2i2}\lambda'^*_{2i3} $ due to the squark exchange in
$B\to K^{(*)}\mu^+\mu^-$ decays. The $\varpi$ denotes $\hat{s}$, magenta
``$\times$" denotes the SM prediction within $1\sigma$ error ranges
of the input parameters, olive solid line denotes the central value
of the SM prediction, and blue ``$\mid$" denotes the SUSY
prediction. The
same goes for Figs. \ref{fig:AFBVK}, \ref{fig:ullps}, \ref{fig:MISemi}, and
\ref{fig:MIuSemi}. }\label{fig:ulplps}
\end{figure}

\begin{figure}[htb]
\begin{center}
\includegraphics[scale=0.6]{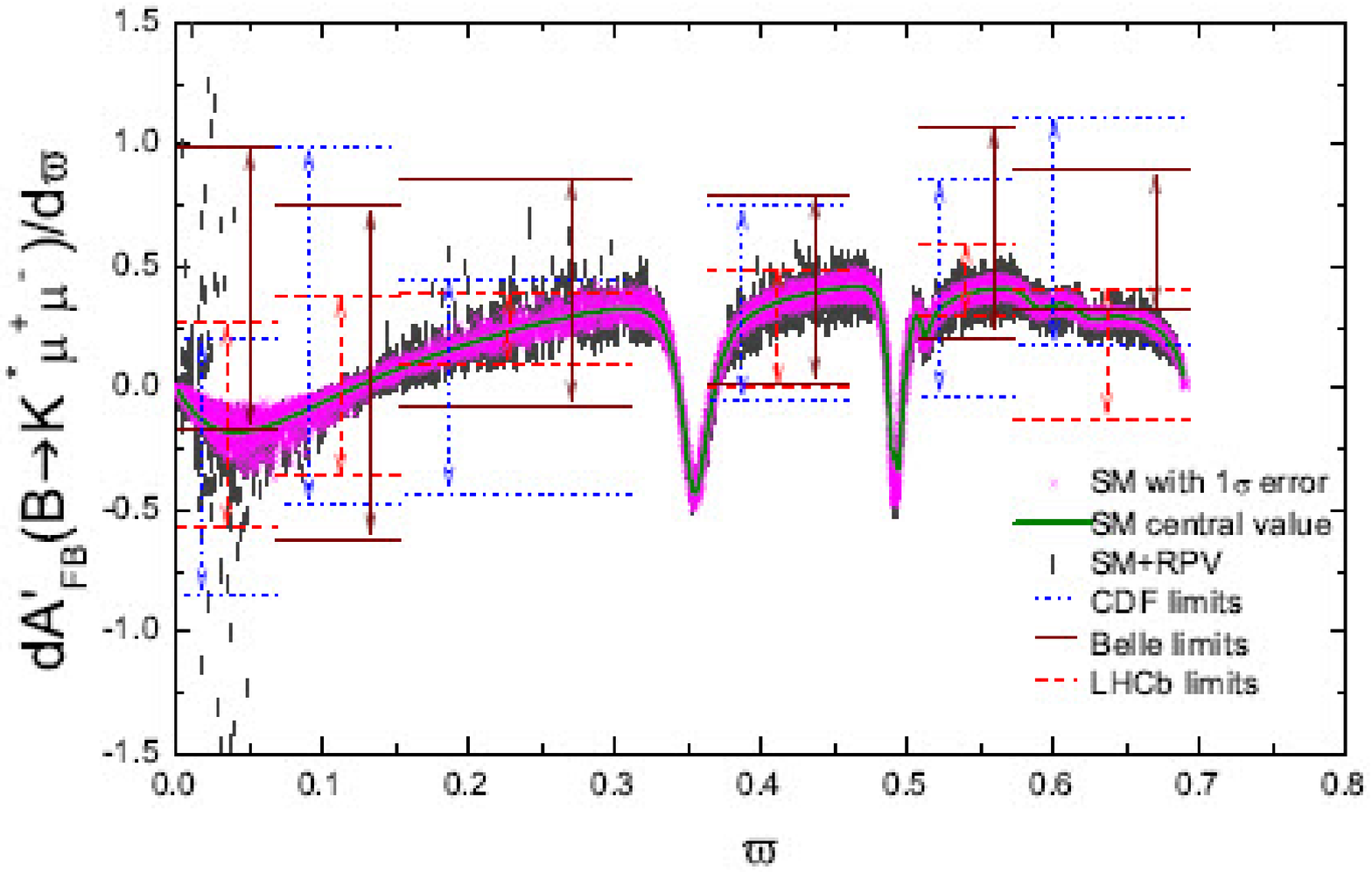}
\end{center}
\vspace{-0.4cm}
 \caption{ $\mathcal{A}_{FB}(B\to K^{*}\mu^+\mu^-)$
including RPV coupling $\lambda'_{2i2}\lambda'^*_{2i3}$ versus the
95\% CL data: CDF (blue dotted line), Belle (purple solid line), and
LHCb (red dashed line).}\label{fig:AFBVK}
\end{figure}
In Fig. \ref{fig:ulplps}, we present correlations between  the
dimuon invariant mass spectra as well as the dimuon forward-backward
asymmetries and the parameter spaces of $\lambda'_{2i3}
 \lambda'^*_{2i2}$ by the two-dimensional scatter
plots. The dimuon invariant mass distribution and the dimuon
forward-backward asymmetry are given with vector meson dominance
contribution excluded in terms of $d\mathcal{B}/d\hat{s}$ and
$d\mathcal{A}_{FB}/d\hat{s}$,  and
 included in  $d\mathcal{B}'/d\hat{s}$ and
 $d\mathcal{A}'_{FB}/d\hat{s}$, respectively.
 In Fig. \ref{fig:ulplps}, the magenta ``$\times$" denotes the SM
prediction within $1\sigma$ error ranges of the input parameters,
olive solid line denotes the central value of the SM prediction, and
blue ``$\mid$" denotes the RPV supersymmetry (SUSY) prediction including
$\lambda'_{2i2}\lambda'^*_{2i3}$ coupling within $1\sigma$ error
ranges of the input parameters. The theoretical uncertainties of the
SM predictions of $d\mathcal{B}(B\to K^{(*)}\mu^+\mu^-)/d\hat{s}$
are quite large; nevertheless the theoretical uncertainties are
canceled a lot in $d\mathcal{A}_{FB}(B\to
K^{*}\mu^+\mu^-)/d\hat{s}$.

The RPV effects on $d\mathcal{A}'_{FB}(B\to
K^{*}\mu^+\mu^-)/d\hat{s}$ are shown in Fig. \ref{fig:ulplps} (f).
This observable has been measured as a function of the dimuon
invariant mass square $q^2$ by  BABAR  \cite{:2008ju}, Belle
 \cite{:2009zv}, CDF  \cite{Aaltonen:2011ja}, and LHCb \cite{Aaij:2011aa}, and the current
situation is specially exemplified in Fig. \ref{fig:AFBVK}. As shown
in Fig. \ref{fig:AFBVK}, the fitted $d\mathcal{A}'_{FB}(B\to
K^{*}\mu^+\mu^-)/d\hat{s}$ from
 Belle is generally higher than the SM expectation
 in whole $q^2$ bins,  and the CDF  fitted result is consistent with the SM prediction in
 some $q^2$ bins and it is higher than the SM prediction in some other $q^2$
 bins;  nevertheless the LHCb fitted result, which is the most precise to data,  is in good agreement with the SM
 prediction.
 Especially, in the region of $0\leq \hat{s}\leq0.072$ (i.e., $0$ GeV$^2\leq q^2\leq2$ GeV$^2$), the Belle measurement
 favors a positive value which is not confirmed by CDF and LHCb, whereas the sign of the SM prediction for $d\mathcal{A}'_{FB}(B\to K^{*}\mu^+\mu^-)/d\hat{s}$
 is negative.
One could find that the constrained RPV coupling $\lambda'_{2i3}
 \lambda'^*_{2i2}$  still could accommodate  $d\mathcal{A}_{FB}(B\to
 K^{*}\mu^+\mu^-)/d\hat{s}$ from Belle, CDF, and LHCb
 at all $\hat{s}$ regions.

As for the  s-channel sneutrino exchange couplings
$\lambda_{i22}\lambda'^*_{i23}$ and $\lambda^*_{i22}\lambda'_{i32}$,
the constraints from $\mathcal{B}(B\to\mu^+\mu^-)$ are rather restrictive.
\begin{figure}[t]
\begin{center}
\includegraphics[scale=0.76]{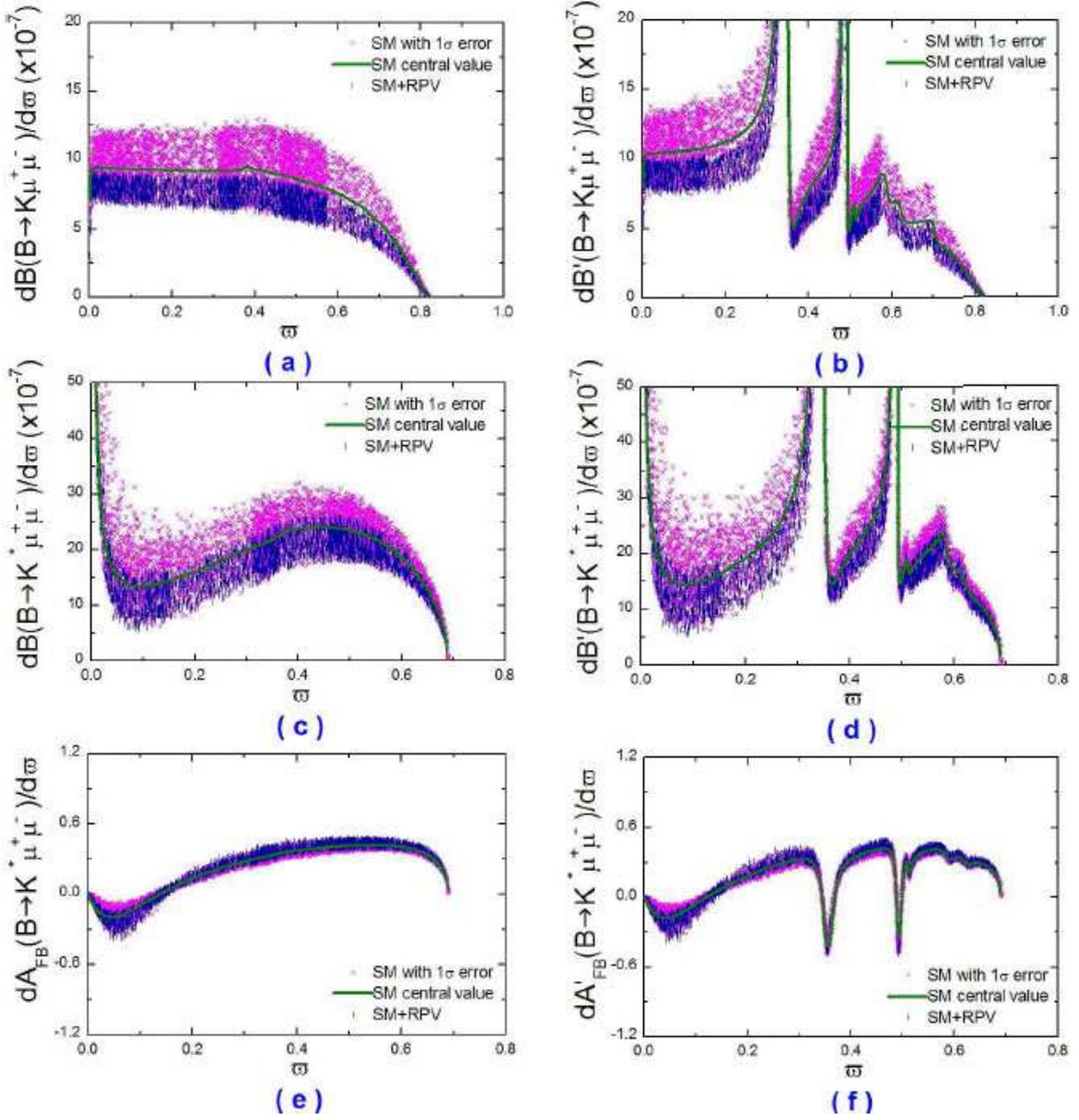}
\end{center}
\vspace{-0.4cm}
 \caption{ The effects of RPV coupling
$\lambda_{i22}\lambda'^*_{i23}$ due to the sneutrino exchange in
$B\to K^{(*)}\mu^+\mu^-$.}\label{fig:ullps}
\end{figure}
The $\lambda_{i22}\lambda'^*_{i23}$ coupling effects in $B\to
K^{(*)}\mu^+\mu^-$ are displayed in Fig. \ref{fig:ullps};  we see
that $\lambda_{i22}\lambda'^*_{i23}$ coupling has negligible
contribution to $d\mathcal{B}(B\to K^{(*)}\mu^+\mu^-)/d\hat{s}$, and
the differences between the SUSY prediction and the SM ones are due
to the 95\% CL experimental constraints. Nevertheless,
constrained $\lambda_{i22}\lambda'^*_{i23}$ coupling has some
effects on  $d\mathcal{A}_{FB}(B\to K^{*}\mu^+\mu^-)/d\hat{s}$.
$\lambda^*_{i22}\lambda'_{i32}$ coupling effects in $B\to
K^{(*)}\mu^+\mu^-$ are similar to $\lambda_{i22}\lambda'^*_{i23}$
effects; thus we will not show them again.

\subsection{RPC MI effects}

Now we study RPC MI effects  in $B_s\to \mu^+\mu^-$ and $B\to
K^{(*)}\mu^+\mu^-$ decays in the MSSM with large tan$\beta$. The
eight kinds of MIs $(\delta^{u,d}_{AB})_{23}$ with $(A,B)=(L,R)$
contribute to $B\to K^{(*)}\mu^+\mu^-$ decays, but only three kinds
of MIs $(\delta^u_{LL})_{23}$, $(\delta^d_{LL})_{23}$, and
$(\delta^d_{RR})_{23}$ contribute to $B_s\to \mu^+\mu^-$ decay.
We will only consider the contributions of $(\delta^u_{LL})_{23}$,
$(\delta^d_{LL})_{23}$, and $(\delta^d_{RR})_{23}$ MIs to $B_s\to
\mu^+\mu^-$ and $B\to K^{(*)}\mu^+\mu^-$ decays in this work.
We take the best-fit values of the constrained MSSM parameters from
the LHC SUSY search results \cite{Heinemeyer:2012dc}:
$m_0=450~GeV,m_{1/2}=780~GeV,A_0=-1110,$ $\mbox{sign}(\mu)>0$, and
$\mbox{tan}\beta=41$. The experimental data shown in Eq.
(\ref{Eq:exp}) will be used to constrain the three kinds of MI
parameters.

MI coupling $(\delta^u_{LL})_{23}$  has some effects on
$\mathcal{B}(B_s\to \mu^+\mu^-)$ and $\mathcal{B}(B\to
K^{(*)}\mu^+\mu^-)$, and the bound of $(\delta^u_{LL})_{23}$ is
 obtained from both $\mathcal{B}(B_s\to \mu^+\mu^-)$ and $\mathcal{B}(B\to K^{*}\mu^+\mu^-)$.
However, for $(\delta^d_{LL})_{23}$ and $(\delta^d_{RR})_{23}$ MI parameters,
the constraints by $\mathcal{B}(B\to K^{(*)}\mu^+\mu^-)$ are rather weak,
which  are mainly derived from $\mathcal{B}(B_s\to \mu^+\mu^-)$.
\begin{figure}[t]
\begin{center}
\includegraphics[scale=0.78]{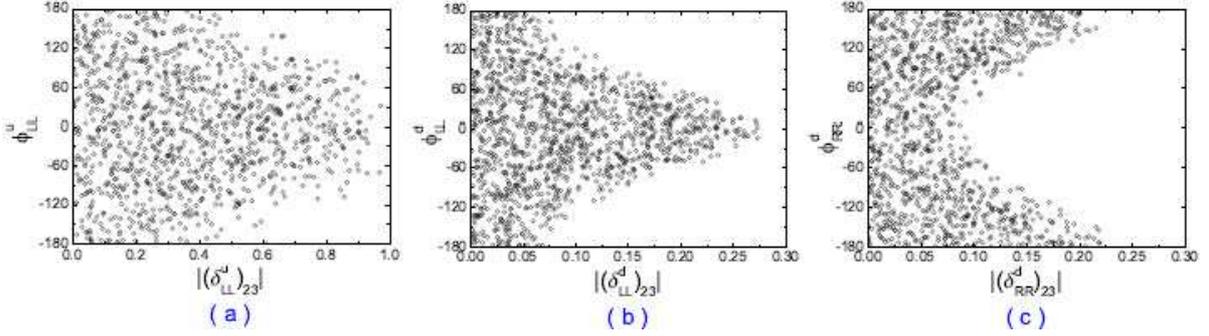}
\end{center}
\vspace{-0.4cm} \caption{ The allowed parameter spaces of
$(\delta^u_{LL})_{23}$,  $(\delta^d_{LL})_{23}$, and
$(\delta^d_{RR})_{23}$ MI parameters constrained by
$\mathcal{B}(B_s\to \mu^+\mu^-)$ and $\mathcal{B}(B\to K^{(*)}
\mu^+\mu^-)$ at 95\% CL,
 and the RPC phases are given in degree. }
\label{fig:boundLLRR}
\end{figure}
The constrained spaces of $(\delta^u_{LL})_{23}$,
$(\delta^d_{LL})_{23}$, and $(\delta^d_{RR})_{23}$ are displayed in
Fig. \ref{fig:boundLLRR}. As shown in Fig. \ref{fig:boundLLRR}, both
phases and moduli of three MIs are obviously constrained by the
branching ratios given in Eq. (\ref{Eq:exp}), and the bounds on the
three moduli are $|(\delta^u_{LL})_{23}|\leq1.0$,
$|(\delta^d_{LL})_{23}|\leq0.28$, and
$|(\delta^d_{RR})_{23}|\leq0.22$. Note that the very strong
constraints on the phases of $(\delta^d_{LL,RR})_{23}$ MIs arise
from $\Delta M_s$, $\Delta \Gamma_s$, and $\phi_s^{J/\psi\phi}$
 \cite{Wang:2011ax}, which are about $\phi^d_{LL,RR}\in
[20^\circ,80^\circ]\cup[-160^\circ,-100^\circ]$ with
$m^2_{\tilde{g}}/m^2_{\tilde{q}}=1$. If considering the strong
constrained phases from $\Delta M_s$, $\Delta \Gamma_s$, and
$\phi_s^{J/\psi\phi}$, we have $|(\delta^d_{LL})_{23}|\leq0.24$ and
$|(\delta^d_{RR})_{23}|\leq0.22$.

\begin{figure}[t]
\begin{center}
\includegraphics[scale=0.6]{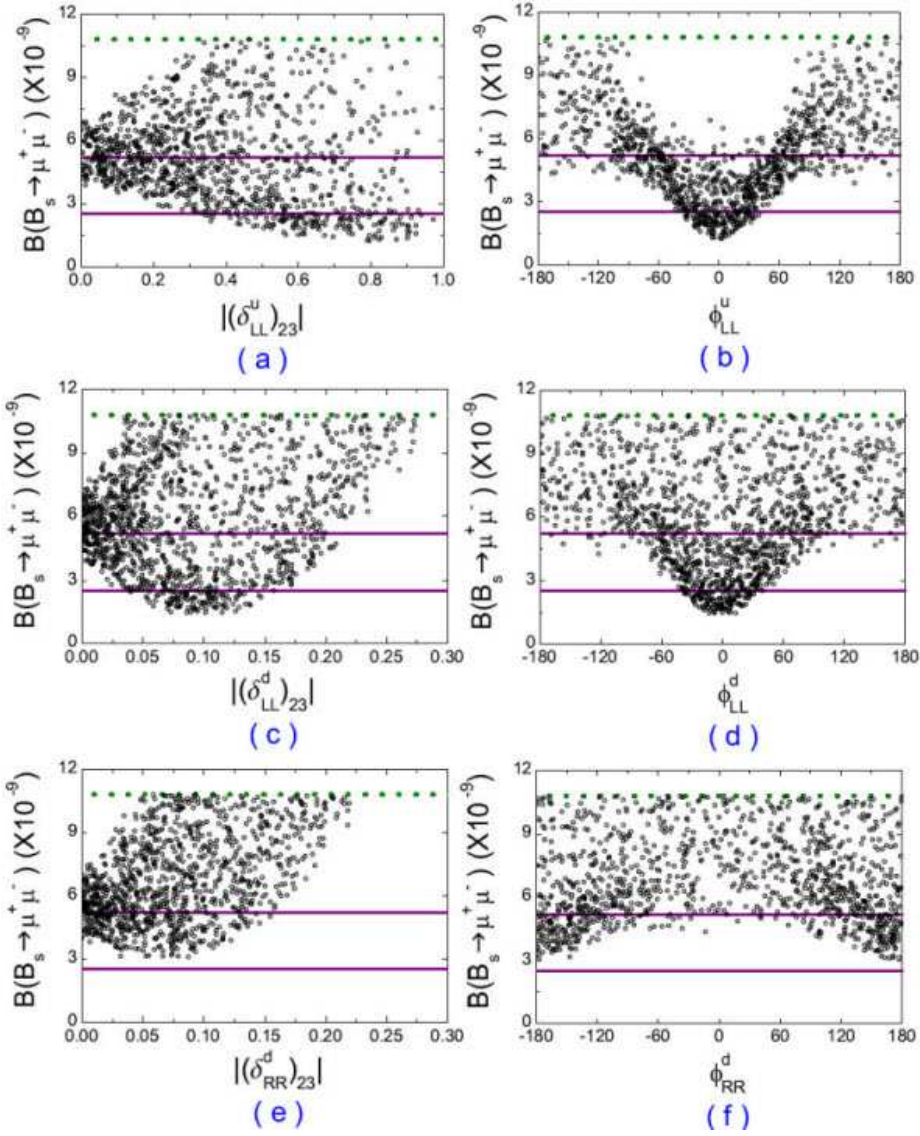}
\end{center}\vspace{-0.4cm}
\caption{ The constrained  MI effects on $\mathcal{B}(B_s\to
\mu^+\mu^-)$. The olive (violet) horizontal dotted (solid) lines
denote the limits of the 95\% CL measurements (SM predictions with
$1\sigma$ error bar).}\label{fig:MIBsmummu}
\end{figure}

Now we analyze the $(\delta^u_{LL})_{23}$, $(\delta^d_{LL})_{23}$,
and $(\delta^d_{RR})_{23}$ MI effects on $\mathcal{B}(B_s\to
\mu^+\mu^-)$. The sensitivities of $\mathcal{B}(B_s\to \mu^+\mu^-)$
to both moduli and phases of three MIs are displayed in Fig.
\ref{fig:MIBsmummu}.
As shown in Fig. \ref{fig:MIBsmummu}, all three couplings are
constrained by the upper limit of $\mathcal{B}(B_s\to \mu^+\mu^-)$,
and $\mathcal{B}(B_s\to \mu^+\mu^-)$ has moderate sensitivities  to
both the moduli and phases. The minimum value of $\mathcal{B}(B_s\to
\mu^+\mu^-)$ may present when $|(\delta^u_{LL})_{23}|\geq0.4$ and
$|\phi^d_{LL}|\leq45^\circ$, $|(\delta^d_{LL})_{23}|\in[0.05,0.15]$
and $|\phi^d_{LL}|\leq45^\circ$ or
$|(\delta^d_{RR})_{23}|\in[0.02,0.10]$ and
$|\phi^d_{RR}|\geq120^\circ$. The differences between the SUSY
predictions at $|(\delta^{u,d}_{AB})_{23}|=0$ and the SM predictions
come from contributions in the MSSM with the CKM matrix as the only
source of flavor violation.

\begin{figure}[t]
\begin{center}
\includegraphics[scale=0.76]{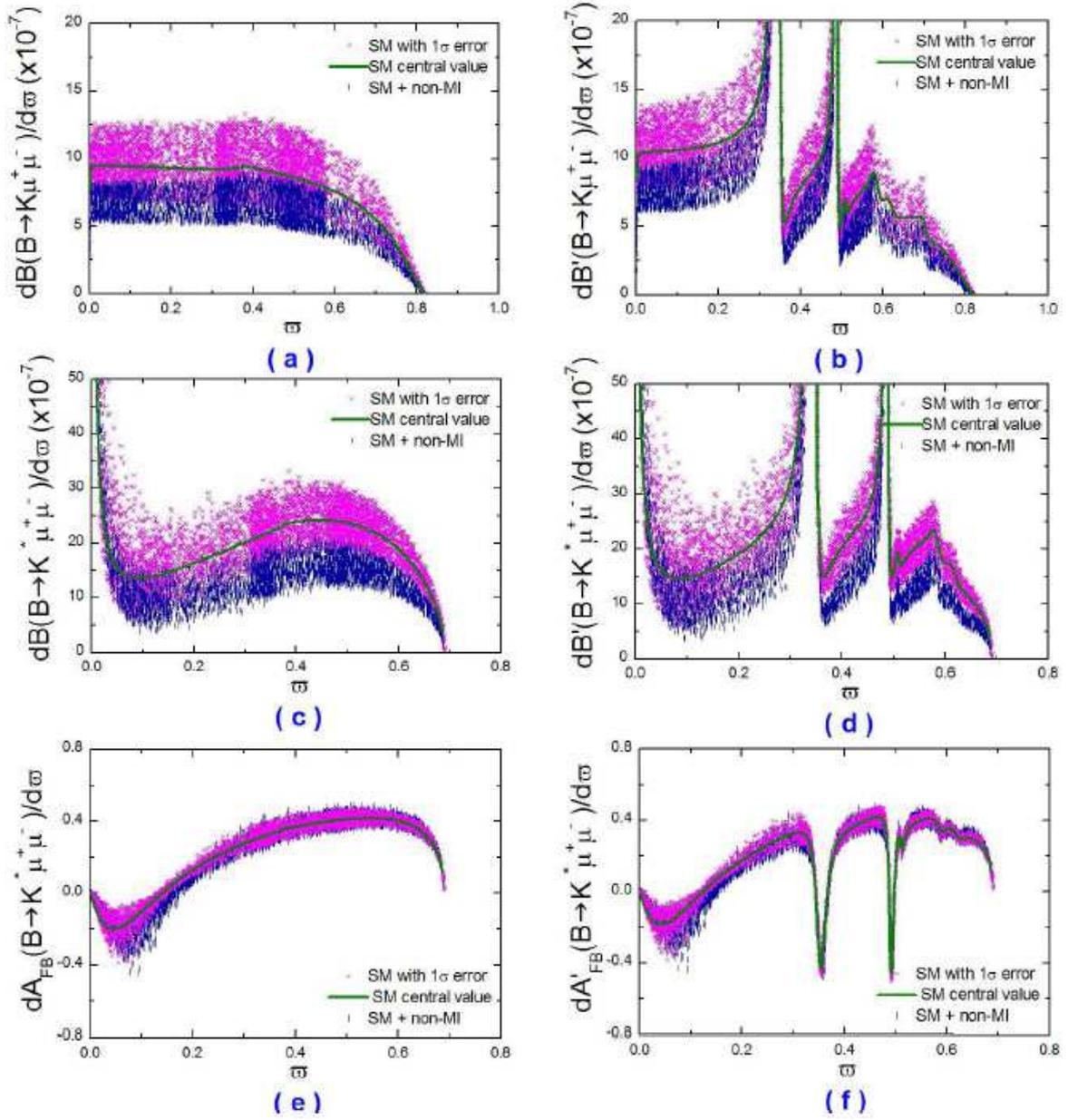}
\end{center}\vspace{-0.4cm}
\caption{ The constrained non-MI effects in $B\to K^{(*)}\mu^+\mu^-$
decays. }\label{fig:MISemi}
\end{figure}

Then we analyze the constrained $(\delta^u_{LL})_{23}$,
$(\delta^d_{LL})_{23}$, and $(\delta^d_{RR})_{23}$ MI effects in
$B\to K^{(*)}\mu^+\mu^-$ decays. Using the constrained parameter
spaces shown in Fig. \ref{fig:boundLLRR}, we will give the MSSM
predictions to the dimuon invariant mass spectra of the decay width
and  the dimuon forward-backward asymmetries of $B\to
K^{(*)}\mu^+\mu^-$ decays in the MI approximation. Besides the MI
contributions,  the SUSY predictions  also include the contributions that
come from graphs including SUSY Higgs bosons and sparticles in the
limit in which we neglect all the MI contributions, which are called
non-MI contributions, and the non-MI SUSY effects are shown in Fig.
\ref{fig:MISemi}.
From Figs. \ref{fig:MISemi} (a-b), we can see that
$d\mathcal{B}(B\to K\mu^+\mu^-)/d\hat{s}$ could be slightly
suppressed  at all $\hat{s}$ regions by the non-MI SUSY couplings. As
shown in Figs. \ref{fig:MISemi} (c-d), $d\mathcal{B}(B\to
K^*\mu^+\mu^-)/d\hat{s}$ could be decreased a lot at the middle
$\hat{s}$ region by these couplings.
Figs. \ref{fig:MISemi} (e-f) show us that the non-MI SUSY couplings
could slightly suppress $d\mathcal{A}_{FB}(B\to
K^{*}\mu^+\mu^-)/d\hat{s}$ at the middle $\hat{s}$ region.

\begin{figure}[b]
\begin{center}
\includegraphics[scale=0.76]{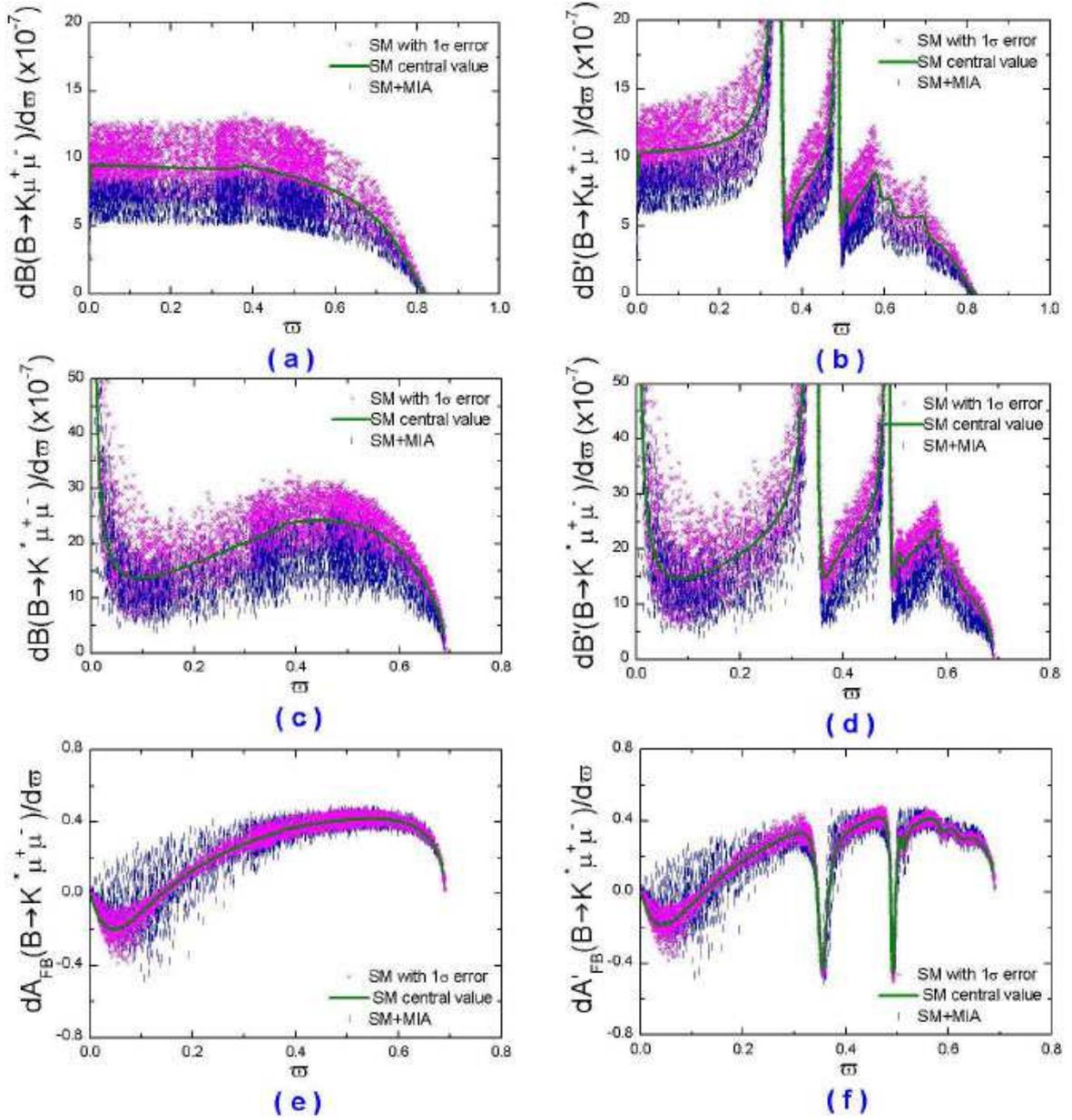}
\end{center}\vspace{-0.4cm}
\caption{ The constrained $(\delta^u_{LL})_{23}$  MI effects in
$B\to K^{(*)}\mu^+\mu^-$ decays. }\label{fig:MIuSemi}
\end{figure}

The constrained $(\delta^d_{LL})_{23}$ and $(\delta^d_{RR})_{23}$
MIs have no obvious effects in $B\to K{(*)}\mu^+\mu^-$ decays.
$(\delta^u_{LL})_{23}$ MI contributions to $B\to K^{(*)}\mu^+\mu^-$
are presented in Fig. \ref{fig:MIuSemi}. Note that the SUSY
predictions in Fig. 8 also include the non-MI contributions shown in
Fig. \ref{fig:MISemi}.
As shown in Figs. \ref{fig:MIuSemi} (a-b),  the constrained
$(\delta^u_{LL})_{23}$ MI has no obvious effects on
$d\mathcal{B}(B\to K\mu^+\mu^-)/d\hat{s}$, which  could be slightly
suppressed at all $\hat{s}$ regions by only non-MI effects. On the
other hand, its contribution to $B\to K^*\mu^+\mu^-$ could be
significant, as shown in Figs. \ref{fig:MIuSemi} (c-f), when
theoretical uncertainties are considered. It is of interest to note
that the contribution to $d\mathcal{A}_{FB}(B\to
K^*\mu^+\mu^-)/d\hat{s}$  is favored
 by the current experimental
measurements from Belle, CDF, and LHCb
\cite{Aaltonen:2011ja,:2009zv,Aaij:2011aa}.

\section{Conclusions}
Motivated by the recent searches of $\mathcal{B}(B_s\to \mu^+\mu^-)$
by the CDF, LHCb, and CMS Collaborations, we have studied  $B_s\to
\mu^+\mu^-$ and $B \to K^{(*)}\mu^+\mu^-$ decays in the MSSM with
and without R-parity.
In the MSSM without R-parity,  we have found that the bounds of
sneutrino exchange RPV couplings are significantly improved by the
present new measurements.  The further constrained RPV coupling due
to t-channel squark exchange still has significant  effects in $B
\to K^{(*)}\mu^+\mu^-$ decays, and the current  measurements of
$d\mathcal{A}_{FB}(B\to K^{*}\mu^+\mu^-)/d\hat{s}$  could be
 accommodated
 by the  squark exchange coupling.
The further constrained couplings due to s-channel sneutrino
exchange could have large  effects in $B_s \to \mu^+\mu^-$, but have
negligible effects in $B \to K^{(*)}\mu^+\mu^-$ decays.

In the MSSM with R-parity, three MI parameters
$(\delta^u_{LL})_{23}$, $(\delta^d_{LL})_{23}$, and
$(\delta^d_{RR})_{23}$ suffer the combined constraints from the
present data of $\mathcal{B}(B_s\to \mu^+\mu^-)$ and $\mathcal{B}(B
\to K^{(*)}\mu^+\mu^-)$. The constrained $(\delta^u_{LL})_{23}$ MI
could give large contributions to $d\mathcal{A}_{FB}(B\to
K^*\mu^+\mu^-)/d\hat{s}$ at all  $\hat{s}$ regions in favor of the
current experimental measurements from Belle, CDF, and LHCb. The
constrained $(\delta^d_{LL,RR})_{23}$ MIs have ignorable effects on
the observables of  $B \to
 K^{(*)}\mu^+\mu^-$decays.
 $d\mathcal{A}_{FB}(B\to K^{*}\mu^+\mu^-)/d\hat{s}$ could be slightly decreased  at the
 middle
$\hat{s}$ region  by the SUSY contributions which come from graphs
including SUSY Higgs bosons and sparticles in the limit in which we
neglect all the MI contributions.

 In the immediate future, the LHC is
expected to become sensitive to $\mathcal{B}(B_s\to \mu^+\mu^-)$.
Accurate measurements of the  $B_s\to \mu^+\mu^-$  and $B \to
K^{(*)}\mu^+\mu^-$ decays could further shrink or reveal the
parameter spaces of MSSM with and without R-parity.

\section*{Acknowledgments}
The work is supported  by the National Science Foundation (Nos.
11105115, 11147136, and 11075059) and the Project of Basic and Advanced,
Technology Research of Henan Province  (No. 112300410021).

\begin{appendix}
\section*{Appendix: Input parameters}
\label{Appendix} The input parameters are summarized in Table
\ref{INPUT}. For the RPC MI effects, we take the five free
parameters $m_0=450~GeV,m_{1/2}=780~GeV,A_0=-1110,$
$\mbox{sign}(\mu)>0$ and $\mbox{tan}\beta=41$ from Ref.
\cite{Heinemeyer:2012dc}.  All other MSSM parameters are then
determined according to the constrained MSSM scenario as implemented
in the program package SUSPECT \cite{suspect2}.
 For the form factors involving the $B\to K^{(*)}$
transitions, we will use the recent  light-cone  QCD sum rules
results  \cite{Ball:2004ye,Ball:2004rg},
 which are renewed with  radiative corrections to
the leading twist wave functions and SU(3) breaking effects.
 For the $q^2$ dependence of the form factors,
they can be parameterized in terms of simple formulas with two or
three parameters. The expression can be found in Refs.
 \cite{Ball:2004ye,Ball:2004rg}.  In our numerical data analysis, the
uncertainties induced by $F(0)$  are also considered.

\begin{table}[ht]
\caption{\small Default values of the input parameters.}
\vspace{0.3cm}\label{INPUT}
\begin{center}
\begin{tabular}{lc}\hline\hline
$m_{B_s}=5.370~GeV,~~m_{B_d}=5.279~GeV,~~m_{B_u}=5.279~GeV,~~m_W=80.425~GeV,$& \\
$m_{K^\pm}=0.494~GeV,~~m_{K^0}=0.498~GeV,~~m_{K^{*\pm}}=0.892~GeV,~~m_{K^{*0}}=0.896~GeV,$& \\
$\overline{m}_b(\overline{m}_b)=(4.19^{+0.18}_{-0.06})~GeV,~~\overline{m}_s(2GeV)=(0.100^{+0.030}_{-0.020})~GeV,$& \\
$\overline{m}_u(2GeV)=0.0017\sim
0.0031~GeV,~\overline{m}_d(2GeV)=0.0041\sim
0.0057~GeV,$& \\
$m_e=0.511\times10^{-3}~GeV,~~m_\mu=0.106~GeV,$~~
$m_{t,pole}=172.9\pm1.1~GeV. $&  \cite{PDG}\\ \hline
$\tau_{B_s}=(1.466\pm0.059)~ps,~~\tau_{B_{d}}=(1.530\pm
0.009)~ps,~~\tau_{B_{u}}=(1.638\pm 0.011)~ps.$&  \cite{PDG}\\\hline
$|V_{tb}|\approx0.99910,~~|V_{ts}|=0.04161^{+0.00012}_{-0.00078}.$&  \cite{PDG}\\
\hline $\mbox{sin}^2\theta_W=0.22306,~~\alpha_e=1/137.$&
 \cite{PDG}\\\hline $f_{B_s}=0.230\pm0.030~GeV.$&
 \cite{Hashimoto:2004hn}\\\hline\hline
\end{tabular}
\end{center}
\end{table}
\end{appendix}

\end{document}